\documentclass{aa}
\usepackage{graphicx}

\newcommand{\etal}{et al.}
\newcommand{\xmm}{{\it XMM-Newton}}
\newcommand{\rxte}{{\it RXTE}}
\def\simlt{\lower.5ex\hbox{\ltsima}}            
\def\simgt{\lower.5ex\hbox{\gtsima}}            

\def\la{~\raise.5ex\hbox{$<$}\kern-.8em\lower 1mm\hbox{$\sim$}~}
\def\ma{~\raise.5ex\hbox{$>$}\kern-.8em\lower 1mm\hbox{$\sim$}~}

\begin{document}

\title{A correlation between the spectral and timing properties of AGN}

\author{I. E. Papadakis\inst{1,2}, M. Sobolewska\inst{2}, P. Arevalo\inst{3}, 
A. Markowitz\inst{4}, I. M. M$^{\rm c}$Hardy\inst{3}, 
L. Miller\inst{5}, J. N.
Reeves\inst{6}, and T.J. Turner\inst{7,8}}

\offprints{I. E. Papadakis;  e-mail: jhep@physics.uoc.gr}
\institute{Physics Department, University of Crete, P.O. Box 2208,
   710 03 Heraklion, Crete, Greece
\and  IESL, Foundation for Research and Technology, 711 10 Heraklion, Greece
\and School of Physics and Astronomy, University of Southampton, Southampton
S017 1BJ, UK
\and Centre for Astrophysics and Space Sciences, University of California, San
Diego, Mail Code 0424, La Jolla, CA 92093-0424, USA
\and Dept. of Physics, University of Oxford, Denys Wilkinson Building,  Keble
Road, Oxford OX1 3RH, UK
\and Astrophysics Group, School of Physical and Geographical Sciences, Keele
University, Keele, Staffordshire ST5 5BG, UK
\and Department of Physics, University of Maryland Baltimore County, 1000
Hilltop Circle, Baltimore, MD 21250, USA      
\and Code 662, Exploration of the Universe Division, NASA/GSFC, Greenbelt, MD
20771, USA
}
\date{Received ?/ Accepted ?} 
\abstract
{We present the results from a combined study of the average X--ray spectral and
timing properties of 14 nearby AGN.} 
{We investigate whether a ``spectral-timing" AGN correlation exists, similar to
the one observed in Cyg X-1, compare the two correlations, and constrain
possible physical mechanisms responsible for the X--ray emission in compact,
accreting objects.}
{For 11 of the sources in the sample, we used all the available data from the
\rxte\ archive, which were taken until the end of 2006.  There are 7795 \rxte\
observations in total for these AGN, obtained over a period of $\sim 7-11$
years. We extracted their 3--20 keV spectra and fitted them with a simple
power-law model, modified by the presence of a Gaussian line (at 6.4 keV) and
cold absorption, when necessary.  We used the best-fit slopes to construct their
sample distribution function, and we used the median of the distribution, and
the mean of the best-fit slopes, which are above the 80th percentile of the
distributions, to estimate the mean spectral slope of the objects. The latter
estimate is more appropriate in the case when the energy spectra of the sources
are significantly affected by absorption and/or reflection effects. We also used
results from the literature to estimate the average spectral slope of the  three
remaining objects.}
{The AGN average spectral slopes are not correlated either with the
black hole mass or the characteristic frequencies that were detected in the
power spectra. They are  positively correlated, though, with the characteristic
frequency when normalised to the sources black hole mass.  This correlation is
similar to the spectral-timing  correlation that has been observed in Cyg X-1,
but not the same.} 
{The AGN spectral-timing correlation can be explained if we assume that the
accretion rate determines both the average spectral slope and the characteristic
time scales in these  systems. The spectrum should steepen and the
characteristic frequency should increase, proportionally, with increasing 
accretion rate. We also provide a quantitative expression between spectral slope
and accretion rate. Thermal Comptonisation models are broadly consistent with
our result, and can also explain the difference between the spectral-timing
correlations in Cyg X-1 and AGN, but only if the ratio of the soft photons'
luminosity to the power injected to the hot corona is proportionally related to
the accretion rate.}

\titlerunning{AGN spectral and timing properties} 
\authorrunning{Papadakis \etal}
\maketitle
   
\section{Introduction}

The X--ray variability properties of AGN have been extensively studied during
the past twenty years. Significant progress has been achieved in the estimation
of their X--ray power spectral density functions (PSDs), which (among other
things) can be helpful in the search for characteristic time scales  in these
objects. This progress has been made possible  with the combined use of
monitoring \rxte\ light curves (which are up to 5--10 years long in many cases) 
and shorter (1 to a few days long), high signal-to-noise, \xmm\ and {\it
Chandra} light curves. The results have shown that the PSD has a $\sim -2$ power
law shape at high frequencies and then,  below a characteristic ``break -
frequency", $\nu_{\rm bf}$, it flattens to a slope of $\sim -1$ (e.g. Uttley,
M$^{\rm c}$Hardy, \& Papadakis 2002; Markowitz \etal\ 2003, M$^{\rm c}$Hardy
\etal\ 2004). Uttley \& M$^{\rm c}$Hardy (2005) (UM05 hereafter) list $\nu_{\rm
bf}$ estimates for 14 nearby AGN. Using these results, M$^{\rm c}$Hardy \etal\
(2006) (M06 hereafter) demonstrated that the corresponding ``break timescale",
$T_{\rm br}=1/\nu_{\rm bf}$,  increases with increasing black hole mass, M$_{\rm
BH}$, and for a given M$_{\rm BH}$, it decreases with increasing accretion rate,
$\dot{m}_{\rm E}$ (in units of the Eddington limit).

Knowledge of the X-ray properties of the Galactic black hole binaries (GBHs) has
also advanced substantially during the past twenty years. Power spectral 
studies of Cyg X-1 in particular have been advanced considerably. Pottschmidt
\etal\ (2003; hereafter P03) for example have used many \rxte\ observations
between 1998 and 2001 to study the long-term evolution of the PSD. They also
studied the energy spectrum of the source and presented convincing evidence that
its timing and spectral properties are closely linked: the characteristic time
scales become shorter  as the spectrum steepens. Axelsson, Borgonovo \& Larsson
(2006; hereafter A06), using several archival \rxte\ observations, which  cover
all spectral states of the source, detected the same ``spectral-timing
properties" correlation as well. Shaposhnikov and Titarchuk (2006) also found
that photon index and characteristic PSD frequencies are positively correlated
in Cyg X-1.

The question whether AGN vary in a manner similar to that of GBHs is a long
standing one. The availability of the better quality AGN light curves over the
last few years (resulted from intense and extended RXTE monitoring campaigns to
sample variability on very long time scales) has allowed a more quantitative
comparison between AGN, Cyg X-1 and other GBHs (GRS1915+105 for example,  in
M06). In this work we used archival \rxte\ observations of 11 AGN, together with
data from the literature for 3 more objects, to estimate their average spectral
slope and compare it with the characteristic frequencies that have been detected
in their power spectra. We show that a positive  ``average spectral slope -
characteristic frequencies" correlation exists, and  we argue that this
correlation is driven by accretion rate, for a given black hole (BH) mass:
objects with a  higher accretion rate relative to the Eddington limit should
also have a steeper spectrum and a shorter characteristic time scale.  We also
compared the spectral-timing relation we found in AGN with a similar relation
that has been detected in one GBH binary, namely Cyg X-1. The two relations
differ, but by an amount that can be explained if we take  into account the BH
mass difference in AGN and Cyg X-1. Our results support the idea that  AGN and
GBHs vary in the same way.  They also have interesting implications regarding
the nature of the X--ray source in AGN and GBHs.

\section{Sample selection and data reduction.}

For the purposes of this study, we need to study AGNs with known $\nu_{\rm bf}$
and average X--ray slopes. The sample of UM05 is the best choice given the
availability of the $\nu_{\rm bf}$ estimates, and the fact that these objects
have been observed regularly with \rxte, over durations of at least a few years.
This is more than $\sim 1-2$ orders of magnitude longer than the characteristic
time scale of the sources in our sample. Most probably then, we have observed
most of the possible flux and spectral variations that they exhibit. 
Furthermore, the study of the same \rxte\ observations that were used to
estimate the power spectrum of the sources, offers us the possibility to 
determine their underlying spectral index around the same period of the power
spectrum measurements.

\begin{table}
\caption{Black hole mass and timing properties of the AGN in the sample.}
\begin{tabular}{ccccc}
\hline
\hline
\newline
Name & M$_{\rm BH}/10^6$ M$_{\odot}$ & Ref. & $\nu_{\rm br}/10^{-7}$Hz & Ref. \\
\hline
Fairall 9    &  255$\pm 56$           	& P04 & 4.0$^{+2.3}_{-0}$ 	& MA03\\
PG~0804+761  &  693$\pm 83$ 		& P04 & 9.6 			& PA03 \\	  
NGC~3227     &  42$\pm 21$  		& P04 & 200$^{+250}_{-140}$ 	& UM05 \\	     
NGC~3516     &  43$\pm 15$  		& P04 & 20$^{+30}_{-10}$	& MA03 \\ 
NGC~3783     &  30$\pm 5$   		& P04 & 40$^{+37}_{-20}$ 	& MA03 \\	  
NGC~4051     &  1.9$\pm 0.8$ 		& P04 & 5050$^{+1100}_{-3000}$ 	& M04 \\	  
NGC~4151     &  13$\pm 5$    		& P04 & 13$^{+19}_{-4}$ 	& MA03 \\	  
Mrk~766      &  3.5          		& W02 & 6100$^{+3500}_{-2500}$  & V03 \\	  
NGC~4258     &  39$\pm 1$    		& H99 & 0.2$^{+23}_{-0}$ 	& MA05 \\	  
NGC~4395     &  0.05$\pm 0.05$ 		& V05 & 19300$^{+9600}_{-15000}$& V05 \\      
MCG~-6-30-15 &  $4.5\pm 1.5$ 		& M05 & 770$^{+1200}_{-310}$ 	& M05 \\  
NGC~5506     &  88	     		& PA04 & 130$^{+830}_{-72}$ 	& UM05 \\ 	  
NGC~5548     &  67$\pm 3$    		& P04 & 6.3$^{+19}_{-0}$ 	& MA03 \\	  
Ark~564      &  2.6$\pm 0.3$  		& B04 & 23000$^{+5800}_{-6600}$ & PA02 \\	  
\hline 
\end{tabular}
\newline
\newline
\noindent Reference used for M$_{\rm BH}$: B04: Botte \etal\ 2004; H99 -- 
Hernstein \etal\ 1999; M05 -- M$^{\rm c}$Hardy \etal\ 2005; PA04 --  Papadakis
2004; P04 -- Peterson \etal\ 2004;  V05: Vaughan \etal\ 2005; W02 -- Woo \& Urry
2002.  Reference used for $\nu_{\rm br}$: MA03 -- Markowitz \etal\ 2003;  M04 --
M$^{\rm c}$Hardy \etal\ 2004; PA02 --  Papadakis \etal\ 2002; PA03 -- Papadakis,
Reig \& Nandra 2003;  V03 -- Vaughan \& Fabian 2003.

\end{table}

In Table~1 we list the 14 AGN in the UM05 sample,  together with their
M$_{\rm BH}$ and $\nu_{\rm bf}$ estimates.  For 11 of these objects we
considered all the \rxte\ observations that were performed until the end of
2006; all data were taken from the public archive. Table~2 lists the date of the
first and last observation, and the total number of \rxte\ observations that we
have used (second and third column, respectively). For the remaining three
objects, namely NGC~4151, NGC~4258 and NGC~4395 we used data from literature to
determine their average spectral shape, as explained in Sections 3.1, 3.2 and
3.3, below. 

\begin{table} 
\begin{center}
\caption{Summary of the  \rxte\ observations.} 
\begin{tabular}{ccc} 
\hline 
\hline 
\newline
Name & Obs. Date & No. of Obs. \\ 
\hline 
Fairall 9 & 1996-11-03/2003-03-01 & 672 \\ 
PG~0804+761 & 1999-01-24/2004-12-23 & 259 \\ 
NGC~3227  & 1996-11-19/2005-12-04 & 1024 \\
NGC~3516  & 1997-03-16/2006-10-13 & 250 \\ 
NGC~3783  & 1996-01-31/2006-11-08 & 875 \\ 
NGC~4051  & 1996-04-23/2006-10-01 & 1268 \\ 
Mrk~766   & 1997-03-05/2006-11-07 & 221 \\
MCG~-6-30-15 & 1996-03-17/2006-12-24 & 1216 \\ 
NGC~5506  & 1996-03-17/2004-08-08 & 627 \\ 
NGC~5548  & 1996-05-05/2006-11-06 & 866 \\
Ark~564   & 1996-12-23/2003-03-04 & 517 \\ 
\hline                                     
\end{tabular} 
\end{center}
\end{table}

We used data from the Proportional Counter Array (PCA; Jahoda \etal\ 1996) only.
The typical duration of each observation was $\sim 1-2$ ksec. The data were
reduced using {\tt FTOOLS} v.6.3. The PCA data were screened according to the
following criteria: the satellite was out of the South Atlantic Anomaly (SAA)
for at least 30 min, the Earth elevation angle was $\geq 10^{\circ}$, the offset
from the nominal optical position was $\leq 0^{\circ}\!\!.02$, and the parameter
ELECTRON-2 was $\leq 0.1$. Appropriate PCA background files \footnote{We used
the latest background model for the faint objects, {\tt
pca\_bkgd\_cmfaintl7\_eMv20051128.mdl} available from the \rxte\ Guest Observer
Facility}, were used to calculate background model energy spectra in the 3--20
keV band.

\section{Spectral analysis}

We extracted STANDARD-2 mode, 3--20 keV (where the PCA is most sensitive), Layer
1, energy spectra from PCU2 only. After background subtraction, we used the {\tt
XSPEC} v.11.3.2 software package (Arnaud 1996) to fit a simple ``power-law +
Gaussian line"  (to account for the K$\alpha$ iron feature) to the spectrum of
each observation. We used PCA response matrices and effective area curves
created specifically for the individual observations by {\tt pcarsp}. All
spectra were rebinned using {\tt grppha} so that each bin contained more than 15
photons for the $\chi^2$ statistic to be valid. 

The simple ``power-law + Gausian" (hereafter PLG) model fitted well almost  all
of the individual \rxte\ spectra of the sources listed in Table~2. The only
exception is   NGC~5506, which hosts a heavily reddened Narrow Line Seyfert 1
nucleus (Nagar et al. 2002). Its central source is absorbed by neutral matter
with column density of $(3{-}4)\times10^{22}$ cm$^{-2}$ (Lamer, Uttley, \&
M$^{\rm c}$Hardy 2000, Bianchi \etal\ 2003). For this reason we used a ``{\tt
wabs}$\times$PLG" model instead, with N$_{\rm H}$ fixed at $3\times 10^{22}$
cm$^{-2}$.  This model fitted well most of the \rxte\ spectra of this source.

\subsection{NGC~4395}

NGC~4395 is a rather faint X--ray source, with an average $2-10$ keV flux of
$\sim 5\times 10^{-12}$ ergs cm$^{-2}$ s$^{-1}$ (Shih, Iwasawa \& Fabian 2003), 
which has not been observed by \rxte. Shih \etal\ have estimated a  mean
spectral slope, $\bar{\Gamma}_{\rm obs}$, of $1.46\pm 0.02$ from the study of a
$\sim 7$ day long {\it ASCA} observation.  Since $T_{\rm br}\sim 500$ s for this
source (Vaughan \etal\ 2005), we can be reasonably confident that the  {\it
ASCA} observation sampled most of the flux/spectral variations that the source
displays. For this reason we adopted this value as the best estimate for its
average spectral shape.

\subsection{NGC~4258}

The nucleus of NGC~4258 is heavily obscured in the X--ray band. Absorption 
column density estimates of the order of a few$\times 10^{23}$ cm$^{-2}$ have
been reported in the past (Yang \etal\ 2007). The absorption also varies on time
scales of months and years (Fruscione \etal\ 2005). NGC~4258 has been observed
extensively by \rxte. There are 707 observations (obtained until the end of
2006) in the archive. We reduced the data for all of them, and fitted the
resulting spectra with a {\tt wabs}$\times$PLG model, keeping N$_{\rm H}$ fixed
at $10^{23}$ cm$^{-2}$. The model fitted the \rxte\ spectra well. The mean
spectral slope is equal to 2.02. 

Fruscione \etal\ (2005) found that the best-fit $\Gamma$ values  increase with
decreasing instrumental angular resolution. They argued that this is due to the
fact that a small extraction radius can isolate  the central engine from the
surrounding soft nuclear emission. It is unsurprising then that the mean
spectral slope we estimated is similar to the best-fit value of $\sim 2.1$ that
Fiore \etal\ (2001) reported from a study of {\it BeppoSax} data, and
significantly steeper than the values obtained from the {\it Chandra} and \xmm\
data analysis (Fruscione \etal\ 2005).  For this reason, we used their results
from the spectral analysis of 9 {\it Chandra} and\xmm\ observations, which were
performed between May 8 2000 and May 22 2002. The weighted mean of their best
fit spectral slopes is $\bar{\Gamma}_{\rm obs}=1.69$, which we adopt as the
average spectral slope estimate for this source. 

\subsection{NGC 4151}

The X--ray spectrum of NGC~4151 is one of the most complex in AGNs,
characterised by narrow and broad  spectral features from soft to hard X--rays
(e.g. Kraemer \etal\ 2005). The spectrum above 2 keV is affected by absorption
from a neutral absorber, which may even be patchy, and from warm material
photionised by the central continuum (e.g. Beckmann \etal\ 2005, de Rosa \etal\
2007).

NGC~4151 has been observed regularly by \rxte. Data from 506 observations 
performed until the end of 2006 are stored in the archive. A simple PLG  model
yields  $\chi^2_{\rm red}>2$ in 241 out of the 506 spectra. The use of a partial
covering fraction absorption model, ``{\tt pcfabs}$\times$PLG", with the 
covering fraction and N$_{\rm H}$ fixed at 0.55 and 1.8$\times 10^{23}$
cm$^{-2}$ (de Rosa \etal\ 2007) improved the quality of the model fit but there 
are still more than a hundred spectra that the model could not fit well. Given
the complexity of the NGC~4151 spectrum and the low spectral resolution of PCA, 
it is not safe to use a more complex model to fit the \rxte\ data.  Using  the
results of de Rosa \etal\ (2007), based on the spectral study of 8 {\it
BeppoSax} observations from January 1996 to December 2001, we found that  the
weighted mean of the best fit spectral slopes is $\bar{\Gamma}_{\rm obs}=1.61$.

\section{Results}

\subsection{The mean spectral slope}

The top panel in Fig.~1 shows the sample distribution function of the  Mrk 766
best-fit spectral slope values, $\Gamma_{\rm obs}$. A Gaussian fits well this
distribution. This is the case with 5 more sources, namely: NGC~5548,
PG~0804+761, NGC~3516, NGC~5506 and Ark 564. In the middle and bottom panels of
the same figure we plot the $\Gamma_{\rm obs}$  distribution functions in the
case of NGC~3227 and NGC~4051.  An extended tail towards low $\Gamma$ values can
be seen in both cases. In NGC~3227, these low-$\Gamma$ values correspond to the
transient absorption by a gas cloud of column density in late 2000 and early
2001 (Lamer, Uttley \& M$^{\rm c}$Hardy, 2003). As for NGC~4051, it is well
established that it shows rather unusual (among AGN)  low flux states, which
last from weeks to months, during which the X-ray spectrum becomes extremely
hard (Uttley \etal\ 2004, and references therein). Less pronounced tails towards
low $\Gamma$ values were also observed in the spectral slope distribution
functions of Fairall 9, NGC~3783 and MCG~-06-30-15.  

Given the asymmetry in the spectral slope distributions  we observe in some
cases,  we used the median of the distribution as an estimate of the mean
spectral slope, $\bar{\Gamma}_{\rm obs}$, for all sources in our sample.  The
results are listed in the second column of Table~3. The numbers in the
parentheses correspond to the standard deviation of the $\Gamma_{\rm obs}$
distributions. They indicate the scatter of the individual $\Gamma_{\rm obs}$
about their mean, and are representative of the typical range of the spectral
slopes that we observe for each object. In the case of NGC~4258 and NGC~4151,
the standard deviation is corrected for the contribution of the error of the
individual best-fit $\Gamma$ values, which in some cases was quite large. In the
case of NGC~4395, the error listed is the one reported by Shih \etal\ 2003.

\begin{figure} 
\centering
\includegraphics[width=8.0cm]{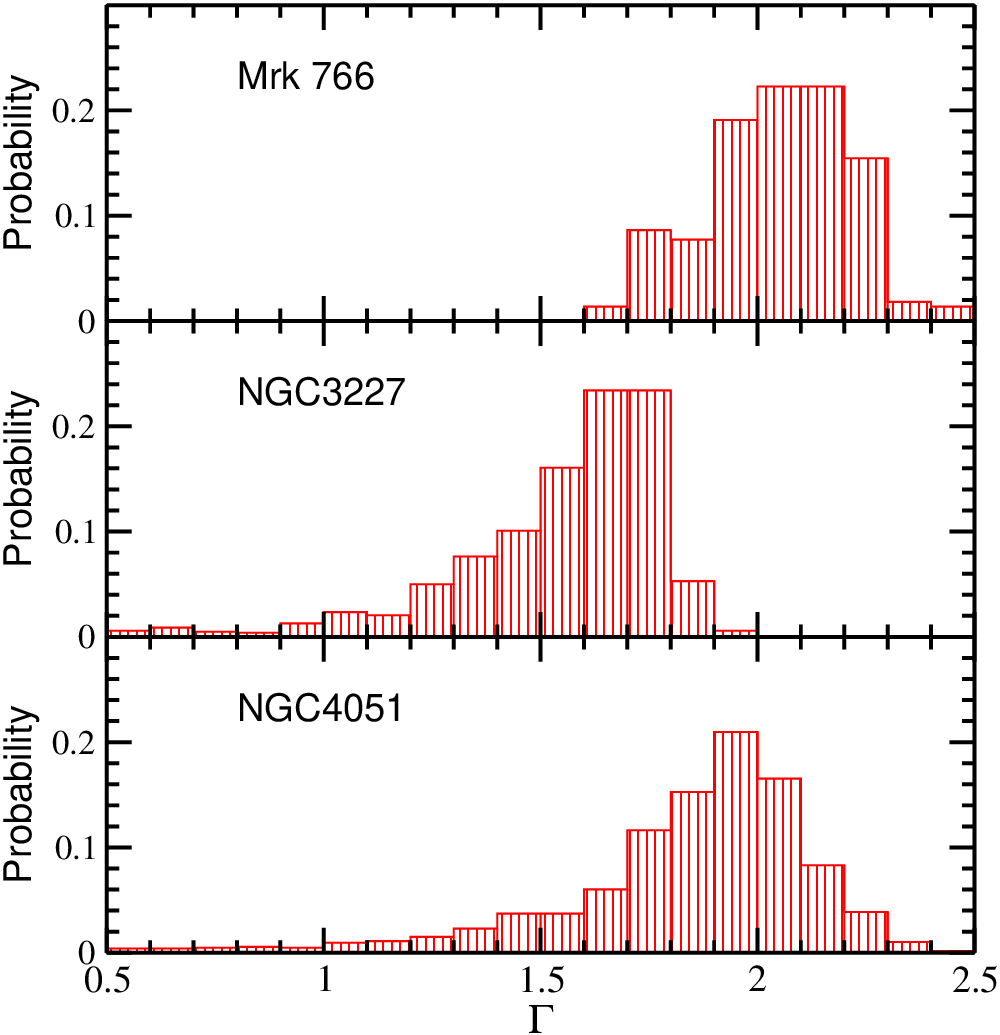} 
\caption{The spectral slope distribution function of Mrk 766 (top
panel), NGC~3227 (middle panel) and NGC~4051 (bottom panel).}
\label{figure:distr} 
\end{figure}

\begin{table}
\caption{The mean spectral slope estimates. }
\begin{tabular}{ccc}
\hline
\hline
\newline
Name & $\bar{\Gamma}_{\rm obs}(\sigma_{\bar{\Gamma}_{\rm obs}})$ 
& $\bar{\Gamma}_{\rm 20\%}(\sigma_{\bar{\Gamma}_{\rm 20\%}})$  \\
\hline
Fairall 9    		& 1.82(0.16) & 2.01(0.08)  \\ 
PG~0804+761  		& 2.03(0.19) & 2.31(0.09)   \\ 	   
NGC~3227     		& 1.61(0.25) & 1.79(0.05)   \\ 	      
NGC~3516     		& 1.49(0.20) & 1.74(0.07)   \\ 	   
NGC~3783     		& 1.66(0.09) & 1.77(0.03)  \\ 	   
NGC~4051     		& 1.90(0.32) & 2.16(0.08)   \\ 	   
NGC~4151$^{\rm a}$     	& 1.61(0.08) & 1.72(0.08)  \\      
Mrk~766      		& 2.06(0.16) & 2.25(0.06)   \\ 	   
NGC~4258$^{\rm b}$     	& 1.69(0.16) & 1.80(0.20)  \\      
NGC~4395$^{\rm c}$     	& 1.46(0.05) &  $-$ \\	 
MCG~-6-30-15 		& 1.86(0.14) & 2.03(0.06)   \\		
NGC~5506     		& 1.85(0.06) & 1.94(0.04)   \\		
NGC~5548     		& 1.75(0.10) & 1.87(0.06)   \\		
Ark~564      		& 2.51(0.22) & 2.82(0.16)   \\		
\hline 
\end{tabular}
\newline
\newline
\noindent Values in the parentheses correspond to the standard deviation of the
points about their mean. \\
\noindent $^{\rm a}$ Values estimated using the results of de Rosa \etal\
(2007).
\newline
\noindent $^{\rm b}$ Values estimated using the results of  Fruscione \etal\
(2005). 
\newline
\noindent $^{\rm c}$ Estimate taken from Shih \etal\ (2003).
\end{table}

\subsection{The ``intrinsic" spectral shape of AGNs}
 
We found that most objects show significant spectral variations (the results
from a detailed analysis of these variations will be presented elsewhere). If
the $\Gamma_{\rm obs}$ variations correspond to variations of the intrinsic
continuum slope, $\Gamma_{\rm int}$,  then $\bar{\Gamma}_{\rm obs}$ will be
representative of the average $\Gamma_{\rm int}$ as well. However,  the AGN
X-ray spectra, in some cases at least, are strongly affected by the presence of
warm absorbing material even at energies above $2$ keV. For example, significant
warm absorbing effects have been observed in  NGC~3516 (Turner \etal\ 2005),
NGC~3783 (Reeves \etal\ 2004), MCG -06-30-15 (Miller, Turner, \& Reeves 2008)
and Mrk 766 (Turner \etal\ 2007).  In this case, $\Gamma_{\rm obs}$ will be a
biased estimate of $\Gamma_{\rm int}$. 

It is even possible that the spectral variations we observed are mainly caused
by variations in a complex and multi-layered absorber, while  $\Gamma_{\rm int}$
remains constant (e.g. Turner \etal\ 2007, Miller \etal\ 2008). If that is the
case, then, for each individual spectrum,  $\Gamma_{\rm obs}=\Gamma_{\rm
int}+\Delta\Gamma$, where $\Delta\Gamma<0$  (as any absorption effects always
result in flatter spectra). Since we expect $\Delta\Gamma$ to be different from
one observation to the other (due to changes in the covering factor, ionization
state of the absorber etc), then $\bar{\Gamma}_{\rm obs}=\Gamma_{\rm
int}+\bar{\Delta\Gamma}$, where  $\bar{\Delta\Gamma}$ is the mean of all the
individual $\Delta\Gamma$s, and should be negative, hence $\bar{\Gamma}_{\rm
obs}<\Gamma_{\rm int}$. 

It has also been suggested that the observed spectral variations are caused by
the combination of a highly variable (in flux) power-law (with $\Gamma_{\rm
int}$=constant) and  a constant reflection component (e.g., Taylor, Uttley \&
M$^{\rm c}$Hardy 2003, Ponti \etal\ 2006, Miniutti \etal\ 2007). Even in this
case, we expect that  $\Gamma_{\rm obs}=\Gamma_{\rm int}+\Delta\Gamma$ (with
$\Delta\Gamma<0$), hence $\bar{\Gamma}_{\rm obs}=\Gamma_{\rm
int}+\bar{\Delta\Gamma}$.

The point is that, if the $2-20$ keV spectrum is affected by absorption and/or
reflection effects, then $\bar{\Gamma}_{\rm obs}$ will be  a biased estimator
of  $\Gamma_{\rm int}$. One way to  minimise the bias is to estimate the mean
spectral slope  using only the largest $\Gamma_{\rm obs}$ for each object since,
in this case,  $\Delta\Gamma$ will be minimum. For this reason, we estimated the
mean of the $\Gamma_{\rm obs}$ which are larger than the 80th percentile of the
distribution (i.e. the value below which 80 percent of the observations fall).
In the case of NGC~4258, we used the mean of the three steepest $\Gamma$ values
of Fruscione \etal\ (2005) as an estimate of $\bar{\Gamma}_{\rm 20\%}$.
Similarly, in the case of NGC~4151 we considered the mean of the two steepest 
$\Gamma$ values of de Rosa \etal\ (2007).

The $\bar{\Gamma}_{\rm 20\%}$ estimates are also listed in  Table~3. The
numbers in the parentheses correspond to  the standard deviation of the points
in the 20\% upper part of the  $\Gamma_{\rm obs}$ distributions, and indicate 
the scatter of these points about $\bar{\Gamma}_{\rm 20\%}$. The 80/20 dividing
line is somehow arbitrary, and is mainly determined by the need to retain a
sizable sample of $\Gamma_{\rm obs}$ values to estimate their mean in the case
of sources with small number of  observations. In any case, our results do not
change significantly when we use the 90th or 70th percentile of the distribution
for the sources with more than 800 observations. Furthermore, the
$\bar{\Gamma}_{\rm 20\%}$ values are closer to spectral slope estimates which
are resulted  when complex, and more realistic,  models are fitted to high
quality X--ray spectra. For example, in  Mrk 766, Miller \etal\ (2007) derive
$\Gamma=2.38\pm 0.04$ from the  ``principal component  analysis" (fit to
eigenvector 1). In NGC 3516, Turner et al. (2005)  derive $\Gamma=1.82\pm 0.01$
and  $1.77\pm 0.02$ from \xmm\ observations at 2 different epochs.  In MCG
-6-30-15, the absorption model in Miller \etal\ (2008)  derives a best-fit of
$\Gamma=2.26\pm 0.02$  (from the {\it Suzaku} data analysis)  and
$\Gamma=2.28\pm 0.01$  from the study of the \xmm\ long looks. In comparison,
Minutti et al. (2007) in their  blurred reflection model also derive a steep
$\Gamma=2.26\pm 0.04$  from the {\it Suzaku} data. Finally,  in NGC 4051, in a
recent paper by Terashima \etal\ (2008; PASJ, submitted), based on absorption
fits to {\it Suzaku} data,  derive $\Gamma=2.04 \pm 0.01$. The average
difference between these spectral slope estimates and $\bar{\Gamma}_{\rm 20\%}$
is less than $\sim 0.1$. Therefore,  $\bar{\Gamma}_{\rm 20\%}$ should be more
representative of the intrinsic spectral slope of each source, if  absorption
and/or reflection effects are significant.

\subsection{The ``spectral-timing" relation in AGN}

Filled squares in the upper panel of  Fig.~2 indicate the ``mean spectral slope
vs characteristic frequency" relation for the AGN in our sample, using the 
$\bar{\Gamma}_{\rm obs}$ values listed in Table~3 and the $\nu_{\rm bf}$
estimates listed in Table~1. The plot in the middle panel shows the ``mean
spectral slope vs BH mass" relation, and in the bottom panel we plot
$\bar{\Gamma}_{\rm obs}$ as a function of the characteristic frequency
multiplied by the BH mass of each object (using the BH mass estimates listed in
Table~1). We call the product $\nu_{\rm bf}\times$M$_{\rm BH}$ as the
``normalised characteristic frequency", $\nu_{\rm norm}$ (in units of
Hz$\times$M$_{\odot}$). The crosses in the bottom panel of Fig.~2 indicate the
``mean spectral slope vs $\nu_{\rm norm}$" relation when we used
$\bar{\Gamma}_{\rm 20\%}$ (instead of $\bar{\Gamma}_{\rm obs}$) as a measure of
the mean spectral slope for the sources in our sample.

Visual inspection of Fig.~2 suggests that, although the mean spectral slope does
not correlate with BH mass, it may correlate positively with $\nu_{\rm bf}$, and
even more so with $\nu_{\rm norm}$: objects with steeper spectra have  shorter
characteristic frequencies as well. We reached the same conclusion even when we
replaced $\bar{\Gamma}_{\rm obs}$ with $\bar{\Gamma}_{\rm 20\%}$ in the first
two panels of Fig.~2. To quantify the correlation of the mean spectral slope
with $\nu_{\rm bf}$, BH mass, and $\nu_{\rm norm}$,  we used the Kendall's
$\tau$ test. The test was performed in the  log-log space, and the results
($\tau$ and P$_{\rm null}$, i.e. the probability that, under the hypothesis there
is no correlation, $\tau$ could be this large or larger just by chance) are
listed in Table~4. We accepted a correlation to be ``significant" if P$_{\rm
null}\le 0.05$. Values in third column of Table~4 are the test results when we
use $\bar{\Gamma}_{\rm 20\%}$ instead of $\bar{\Gamma}_{\rm obs}$.

\begin{figure}   
\centering  
\includegraphics[width=8.0cm]{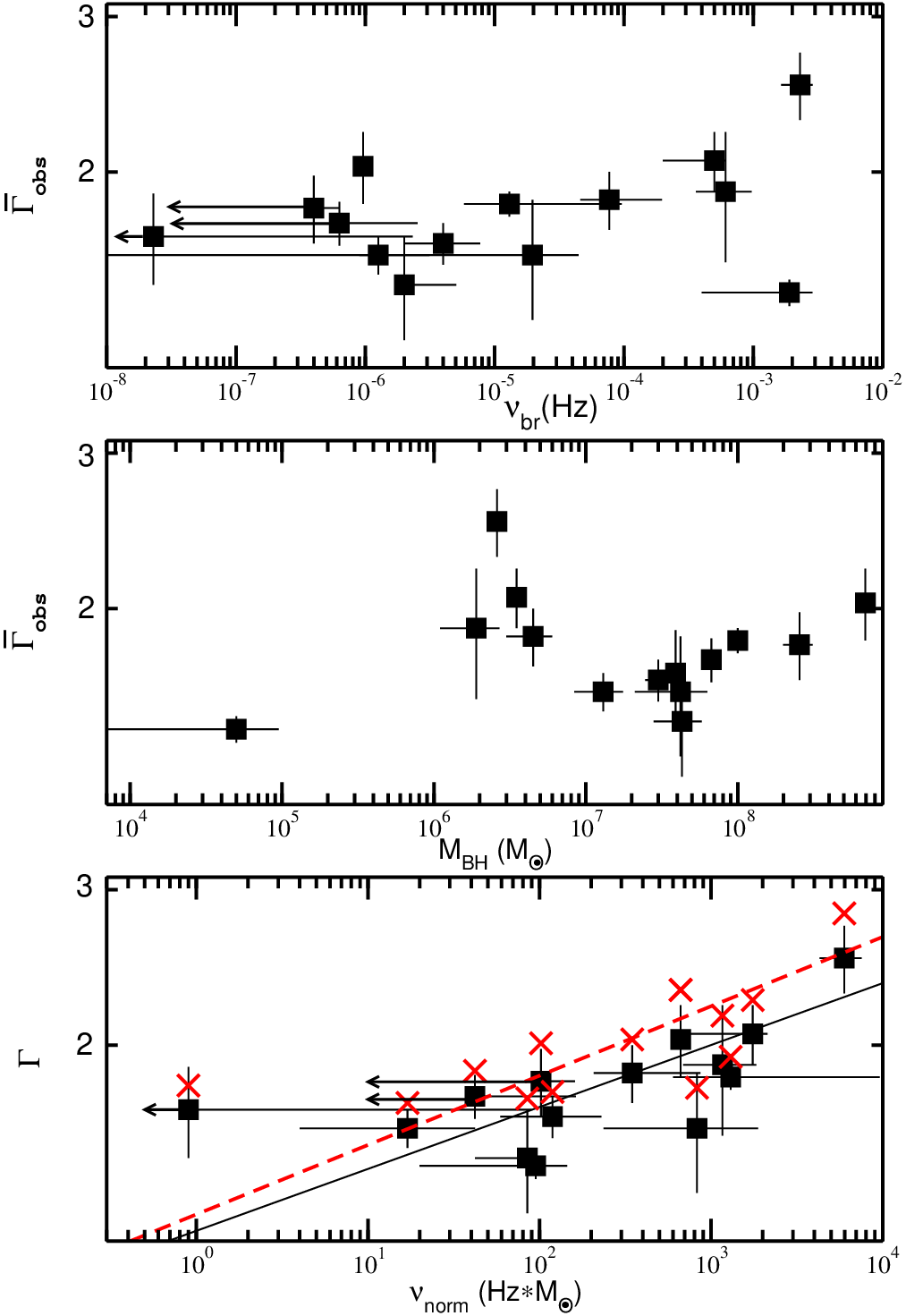} 
\caption{{\it Top panel}:  The ``$\bar{\Gamma}_{\rm obs}$ - break frequency"
relation for the AGN in our sample. {\it Middle panel}:  The
``$\bar{\Gamma}_{\rm obs}$ - BH mass" relation for the same objects. {\it Bottom
panel}: The  AGN ``mean spectral slope - normalised characteristic frequency"
relation. Filled squares and  crosses indicate the($\bar{\Gamma}_{\rm
obs},\nu_{\rm norm}$) and the ($\bar{\Gamma}_{\rm 20\%},\nu_{\rm norm}$) data,
respectively.  The solid and dashed lines  in the same panel indicate the 
best-fit power-law model to the  ($\bar{\Gamma}_{\rm obs}, \nu_{\rm norm}$) and 
($\bar{\Gamma}_{\rm 20\%},\nu_{\rm norm}$) data, respectively.}
\label{figure:res}  
\end{figure}

\begin{table}
\caption{The Kendaull's $\tau$ test results for the correlations plotted in 
Fig.~2.}
\begin{tabular}{ccc}
\hline
\hline
 & $\tau/$P$_{\rm null}^{\rm a}$ & $\tau^{\rm b}$/P$_{\rm null}$  \\
\hline
$\bar{\Gamma}$ vs $\nu_{\rm bf}$   & 0.20/0.32 & 0.21/0.30 \\
$\bar{\Gamma}$ vs BH mass          & 0.11/0.58 & 0.10/0.62 \\
$\bar{\Gamma}$ vs $\nu_{\rm norm}$ & 0.53/$8.2\times 10^{-3}$ 
& 0.54/$7.3\times 10^{-3}$ \\
\hline 
\end{tabular}
\newline
\noindent $^{\rm a}$P$_{\rm null}$ depends on both the size of the sample
and the value of $\tau$. 
\newline
\noindent $^{\rm b}$ Results in the case we use $\bar{\Gamma}_{\rm 20\%}$
instead of $\bar{\Gamma}_{\rm obs}$.
\end{table}

The Kendall's $\tau$ results imply that only the  ``mean spectral slope vs 
$\nu_{\rm norm}$" correlation is strong (i.e $\tau>0.5$) and significant. This 
is true irrespective of whether we use $\bar{\Gamma}_{\rm 20\%}$ or
$\bar{\Gamma}_{\rm obs}$. The correlation remains significant even if we
omit Ark~564 (i.e. the source with the softest spectrum and the highest
characteristic frequency): $\tau=0.45$, P$_{\rm null}=0.03$ for the 
$\bar{\Gamma}_{\rm 20\%}$--$\nu_{\rm norm}$ relation, and   $\tau=0.46$, P$_{\rm
null}=0.03$ when we use $\bar{\Gamma}_{\rm obs}$ instead of  $\bar{\Gamma}_{\rm
20\%}$. We therefore conclude that, although the mean spectral slope does not
 correlate significantly either with power spectrum break frequency or with
the BH mass, it does correlate positively with the break frequency when
normalised to the BH mass of each object. \footnote{We got the same results when
we excluded from the sample the three sources for which the mean spectral slope
estimation was not based on the analysis of a large number of \rxte\
observations. For example we found that $\tau=0.38$ in the case  of the
``$\bar{\Gamma}_{\rm 20\%}$ -- $\nu_{\rm br}$" relation (P$_{\rm null}=0.10)$,
$\tau=-0.13$, and P$_{\rm null}=0.58$ in the case of the ``$\bar{\Gamma}_{\rm
20\%}$ -- M$_{\rm BH}$" relation, and $\tau=0.53$, P$_{\rm null}=0.02$ in the
case of the  ``$\bar{\Gamma}_{\rm 20\%}$ -- $\nu_{\rm norm}$" relation.}

 To investigate whether a straight line or a  power law model fits best  the 
AGN spectral-timing relation, we considered the 10 AGN that have well measured
PSD breaks and we  fitted their ``average spectral slope - normalised frequency"
data  with a straight line in both the linear and log-log space, taking into
account the errors on both the $\nu_{\rm norm}$ and $\bar{\Gamma}_{\rm obs}$ (or
${\Gamma}_{20\%}$) values. We found that the power-law model (i.e. a straight
line in the log-log  space) describes better the observed spectral-timing
relation. For example, the linear model fit to the ($\bar{\Gamma}_{20\%}$,
$\nu_{\rm norm}$) data resulted in a $\chi^{2}$ value of 25.1 for 8 degrees of
freedom (dof). A linear model to the logarithm of the same data set resulted in
17.6 for 8 dof. To quantify if this change in $\chi^2$ is significant, we
computed the  ``ratio of likelihoods", $L_{1}/L_{2}$ (Mushotzky 1982). It is
defined as
$L_{1}/L_{2}$=exp$[(\chi^2_{2}-\chi^{2}_{1})/2]=$exp$(\Delta\chi^2/2)$, where 
$\chi^{2}_{1}$ and $\chi^{2}_{2}$ are the chisquare for the line fits in the
logarithmic and linear spaces, respectively. We found that $L_{1}/L_{2}=42.5$.
This result suggests that a power law is $\sim 40-45$ times more likely, than a
straight line, to be the ``correct" model for the spectral-timing data plotted
in the bottom panel of Fig.~2.

To derive the best-fit power law parameter values, we used the ``ordinary least
squares bisector" method of  Isobe \etal\ (1990) to fit  the
[log($\bar{\Gamma}_{\rm obs})$, log($\nu_{\rm norm}$)] data with a straight line
of the form log($\bar{\Gamma}_{\rm obs})=a+b$log($\nu_{\rm norm}$). The best-fit
parameter values are $a=0.09\pm 0.05$ and $b=0.070\pm0.015$  (the best-fit model
is indicated by the  solid line in the bottom panel of Fig.~2). The dashed line
in Fig.~2 indicates the best-fit straight line model  to the 
[log$(\bar{\Gamma}_{\rm 20\%})$,log($\nu_{\rm norm}$)]  data. The best-fit
parameter values in this case are $a=0.11\pm 0.06$ and $b=0.079\pm 0.019$. In
other words, a power-law model of the form  $\bar{\Gamma}_{\rm obs}=1.23(\pm
0.14)\nu^{0.07\pm 0.015}_{\rm norm}$ fits well the ($\bar{\Gamma}_{\rm obs},
\nu_{\rm norm}$) data, while in the case of the  ($\bar{\Gamma}_{\rm 20\%},
\nu_{\rm norm}$) data the best-fit power-law model is $\bar{\Gamma}_{\rm
20\%}=1.3(\pm 0.2)\nu^{0.079\pm 0.019}_{\rm norm}$. We note that although the
best-fit slopes are small, they are significantly different than zero.  The
best-fit parameter values are consistent within the errors and their weighted
mean value is  $\bar{a}=0.10\pm0.04$ and $\bar{b}=0.073\pm 0.012$. We therefore
conclude that the spectral-timing relation in AGN is well parametrised by a
power-law model of the form:  $\Gamma_{\rm AGN}=1.26(\pm 0.12)\nu^{0.073\pm
0.012}_{\rm norm}$.

According to M06, the PSD break frequency depends on both the BH mass and
accretion rate approximately as follows: $\nu_{\rm bf}\propto \dot{m}_{\rm
E}/$M$_{\rm BH}$. Consequently,  $\nu_{\rm norm}$ should depend on $\dot{m}_{\rm
E}$ only. Given this observational result, the  $\bar{\Gamma}_{\rm AGN}\propto
\nu_{\rm norm}^{0.07}$ relation  we found can be translated to a 
$\bar{\Gamma}_{\rm AGN}\propto \dot{\rm m}_{\rm E}^{0.07}$ relation. In other
words, our results suggest that  the mean  spectral slope in AGN correlates
positively with $\dot{\rm m}_{\rm E}$: objects with higher accretion rate should
also have steeper spectra (on average).

In the case of Comptonisation models, where thermal electrons in a corona above
the disc upscatter soft photons emitted by the disc of temperature T$_{\rm s}$,
the produced X--rays have a power-law spectrum. When a soft photon of initial
energy $\epsilon_s$ is Compton scattered in the corona, it acquires an energy 
$A\epsilon_s$, on average, where $A$ is the Compton amplification factor 
defined as $A=(L_{\rm diss}+L_{\rm s})/L_{\rm s}$ ($L_{\rm diss}$ and $L_{\rm
s}$ are the power used to heat the corona and the intercepted soft luminosity,
respectively). The produced X--rays have a power-law spectrum whose slope,
$\Gamma$, depends on the temperature and optical depth of the corona. In
general, one expects that when $L_{\rm s}$ increases, the corona cooling will be
more efficient. Consequently, its temperature should decrease and the resulting
X--ray  spectrum will be steeper (i.e. $\Gamma$ will increase).  

Using the numerical Comptonisation code of Coppi (1999; {\tt  eqpair} in {\tt
XSPEC}),which is applicable in the case of a soft photon source located in the
centre of a static, isothermal spherical corona, Beloborodov (1999) derives a
relationship between $\Gamma$ and $L_{s}/L_{diss}$ for different disc
temperatures, Ts. For $T_{s}\sim$ few $\times 10^{4}K$, which is applicable in
AGN, he finds that $\Gamma_{\rm AGN} \sim 2.3(L_{s}/L_{diss})^{0.1}$. For low
mass black holes like Cyg X-1, where $T_{s} \sim$few $\times 10^{6}K$,
$\Gamma_{\rm GBHs} \sim 2.3(L_{s}/L_{diss})^{0.17}$. In the case of a
non-static, outflowing corona, Malzac, Beloborodov \& Poutanen (2001) found
similar results:  $\Gamma_{\rm AGN} \sim 2.2(L_{s}/L_{diss})^{0.07}$ and
$\Gamma_{\rm GBHs} \sim 2.2(L_{s}/L_{diss})^{0.13}$.

We found that  $\Gamma_{\rm AGN}\propto \nu_{\rm norm}^{\sim 0.07}$ which, when
combined with the $\nu_{\rm norm}\propto \dot{\rm m}_{\rm E}$ result of M06,
implies that $\Gamma_{\rm AGN}\propto \dot{\rm m}_{\rm E}^{\sim 0.07}$.  
Consequently,  thermal Comptonisation model predictions, i.e 
$\Gamma\propto(L_{s}/L_{diss})^{0.07}$ $^{\rm or }$ $^{0.10}$,  are consistent
with our results, i.e $\Gamma_{\rm AGN}\propto \dot{\rm m}_{\rm E}^{\sim
0.07}$,  but only if $L_{\rm s}/L_{\rm diss}\propto \dot{\rm m}_{\rm E}$.

\section{Comparison with Cyg X-1}

To compare the spectral-timing behaviour of AGN with that of Galactic black hole
X-ray binary systems, we considered the extensive spectral and timing
observations of the best studied GBH, Cyg X-1. To maximise the spectral-timing
range, we considered all available information, covering ``hard" state
observations (from P03 and A06) and ``soft" and ``intermediate" states (from
A06).

It is important that a consistent measure of characteristic break-frequency is
used for all the Cyg X-1 and AGN data. In the soft state the PSD of Cyg X-1 is
well described by a power-law with only one ``bend", with slope $\sim -1$ at low
frequencies and  $\sim -2$ at high frequencies. This shape is sometimes also
parametrised as a power-law of slope -1 with an exponential cut-off (model 5 of
A06). This shape also fits well the PSDs of almost all AGN that have been
studied so far  (except for Akn564, for which a double Lorentzian model fits
best, M$^{\rm c}$Hardy \etal\ 2007). For the  soft state PSDs of Cyg X-1
therefore, we considered from model 5 of A06, the best-fit ``turnover" or
``bend" frequency at which the PSD slope bends from $-1$ to $-2$. 

In the hard state, the PSD of GBHs such as Cyg X-1 can be fitted either as a
doubly bending power-law of slope 0 at the lowest frequencies, slope $-1$ at
intermediate frequencies, and slope ~ $-2$ at the highest frequencies or, where
the signal-to-noise ratio is higher, as the sum of a number of Lorentzian-shaped
components. Both P03 and A06 have opted for the second option, in modeling the
Cyg X-1 PSD in the hard and intermediate state.  To include therefore the hard
state Cyg X-1 PSD observations, we must use the Lorentzian which, in the bending
power-law parametrization, is closest to the frequency at which the slopes
change from $-1$ to $-2$. In the observations of P03, that Lorentzian is
referred to as $\nu_{2}$, and has an average frequency around 2 Hz. In the
analysis of A06 (and also of Axelsson, Borgonovo and Larsson, 2005), the hard
and intermediate state PSD is parametrised by the sum of two Lorentzians but
also with the addition of a weak cut-off power-law.  The Lorentzian with the
higher frequency corresponds to $\nu_{2}$ of P03 and so we used the frequency of
that Lorentzian here.\footnote{We note that if band-limited PSD power is fitted
by Lorentzians, the Lorentzian at the upper band limit will lie at a slightly
lower frequency than the bend frequency, if the same PSD is fitted by a bending
power-law (cf Akn564, M$^{\rm c}$Hardy \etal\ 2007). The difference, however,
is typically only a factor of 2 or less, much less than the range of frequencies
covered here, and so we do not try and take account of it here.}

In order to measure $\Gamma_{\rm obs}$, P03 fit their spectra with a ``power-law
+ multi-temperature disc-black body + reflection" model  from which the
$\Gamma_{\rm obs}$ of the power-law can be taken. A06 list (20-9)/(4-2) keV
hardness ratios (HR). We have used the empirical relationship between $\Gamma$
and HR of Axelsson \etal\ (2005) to convert HR into $\Gamma_{\rm obs}$.

 The resulting Cyg X-1 data are shown in Fig.3. Open and filled circles indicate
the ($\Gamma_{\rm obs},\nu_{2}$) data for Cyg X-1  where $\nu_2$ is the centroid
frequency of the ``second" Lorentzian in the P03 and A06 model fits to the Cyg
X-1 PSD, respectively. These data define the Cyg X-1  spectral-timing relation 
in its hard state. Crosses indicate the ($\Gamma_{\rm obs},\nu_{c}$) data for
Cyg X-1, where $\nu_c$ is the ``bend" frequency of the ``bending" power-law
model which fits best the Cyg X-1 PSD in its soft state, according to A06 (their
model 5). To convert the observed frequencies (both $\nu_2$ and $\nu_c$) to
$\nu_{\rm norm}$ we assumed a black hole mass of 15  M$_{\odot}$, intermediate
between the two published estimates of 10 M$_{\odot}$ (Herrero et al. 1995) and
20 M$_{\odot}$ (Ziolkowski 2005). We can see that the Cyg X-1 hard and soft
state data form a smooth continuous distribution in the spectral-timing plane
shown in Fig.~3. 

Filled squares in the top and bottom panels of Fig.~3 indicate the  ``spectral
slope -- characteristic frequency" relation for the AGN in our sample, when we
use the $\bar{\Gamma}_{\rm obs}$ and $\bar{\Gamma}_{20\%}$ values,
respectively.  Although in both Cyg X-1 and AGN  the characteristic frequencies
increase as the energy spectrum steepens,  Fig.~3 shows clearly that the
respective ``$\Gamma-\nu_{\rm norm}$" relations are not the same.  

This difference can be explained by the fact that the accretion disc is hotter
in Cyg X-1 than in AGN. So, for example, in the case of an outflowing corona, as
stated in Section 4.3, $\Gamma \propto (L_{s}/L_{diss})^{0.07}$ in AGN but
$\Gamma \propto (L_{s}/L_{diss})^{0.13}$ in Cyg X-1. If  $(L_{\rm s}/L_{\rm
diss})\propto \dot{\rm m}_{\rm E}$, both in AGN and Cyg X-1, then we expect
that  $\Gamma_{\rm AGN}\propto \dot{\rm m}_{\rm E}^{0.07}$ and  $\Gamma_{\rm Cyg
X-1}\propto \dot{\rm m}_{\rm E}^{0.13}$. Using the  $\nu_{\rm norm}\propto
\dot{\rm m}_{\rm E}$ relation of M06, these relations can be written as follows:
$\Gamma_{\rm AGN}\propto \nu_{\rm norm}^{0.07}$ and  $\Gamma_{\rm Cyg
X-1}\propto \nu_{\rm norm}^{0.13}$.  As a result, in the case when $\Gamma_{\rm
Cyg X-1} \approx \Gamma_{\rm AGN}$, we should expect that $\nu_{\rm
norm,AGN}^{0.07}\sim  \nu_{\rm norm,Cyg X-1}^{0.13}$, i.e.  $\nu_{\rm norm,Cyg
X-1}\sim  \nu_{\rm norm,AGN}^{0.55}$ (in the case of a static corona, we expect
a similar relation, i.e.  $\nu_{\rm norm,Cyg X-1}\sim  \nu_{\rm
norm,AGN}^{0.6}$). Open squares in both panels of Fig.~2 indicate  the AGN data
when we transform $\nu_{\rm norm,AGN}$ to  $\nu_{\rm norm,AGN}^{0.55}$. The
agreement now between the Cyg X-1 and AGN ``spectral slope - characteristic
frequency" relation is good (a more detailed derivation of the relation between
$\nu_{\rm norm,AGN}$ and $\nu_{\rm norm,Cyg X-1}$ in the case of thermal
Comptonisation models is presented in Section 6.2).

\begin{figure}  
\centering 
\includegraphics[width=8.0cm]{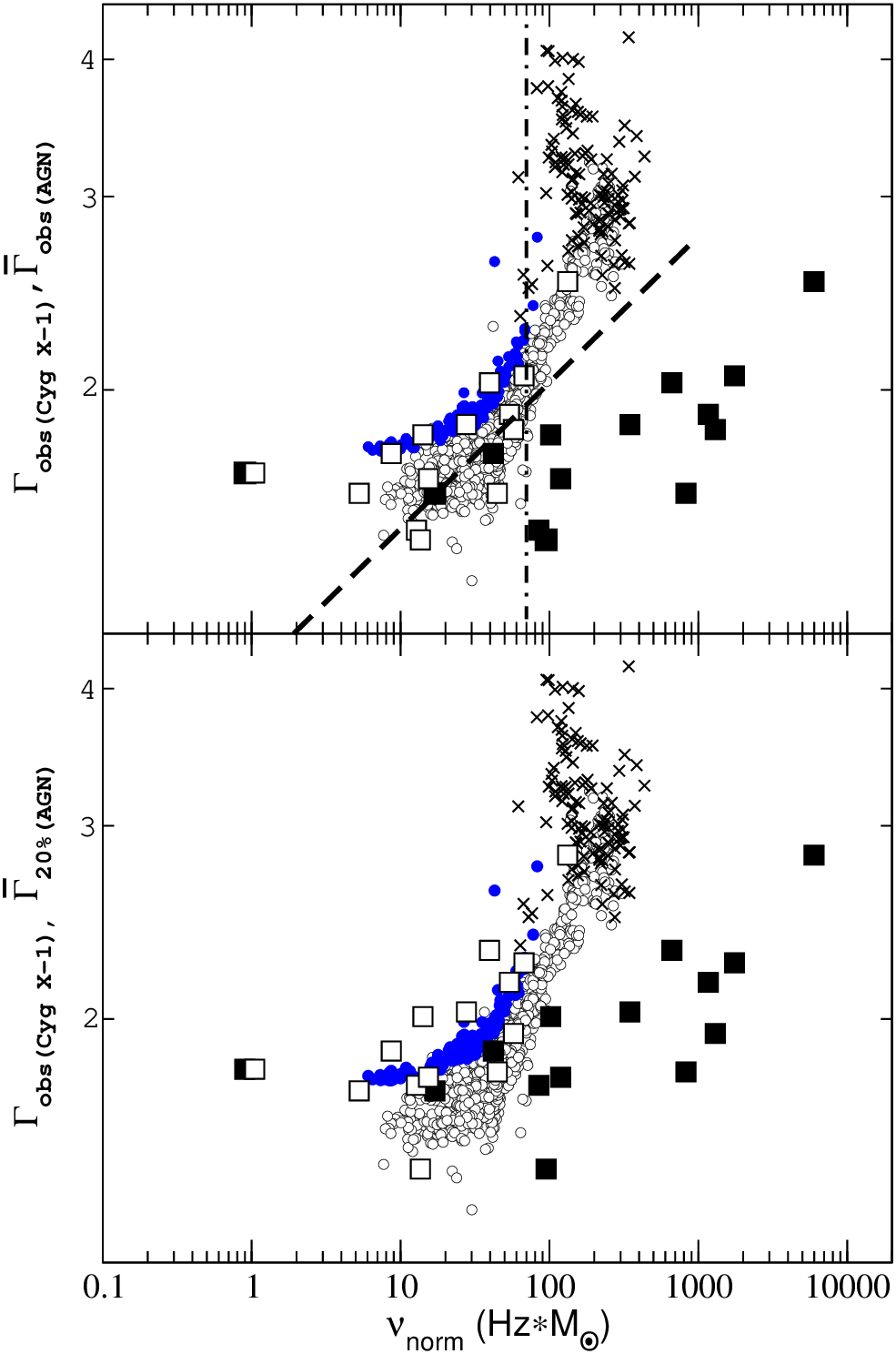} 
\caption{(Both panels)  Filled and open circles indicate the
spectral-timing data for Cyg X-1 in its hard state, using data from P03 and
A06, respectively. Crosses indicate the spectral-timing data for Cyg X-1 in
its soft state using data from A06. Filled squares indicate the ``mean spectral
slope - normalised characteristic frequency" data for the AGN in our sample,
when we consider the $\bar{\Gamma}_{\rm obs}$ and the $\bar{\Gamma}_{20\%}$
values listed in Table~3 (top and bottom panels, respectively). Open squares
indicate  the same AGN data when $\nu_{\rm norm,AGN}$ are shifted to $\nu_{\rm
norm,AGN}^{0.55}$ (see text for details).}
\label{figure:comp}  
\end{figure}

\section{Discussion and conclusions}

We have used 7795 \rxte\ observations of 11 AGN, obtained over a period of $\sim
7-11$ years, to extract their 3--20 keV spectra. We fitted  them with a simple
``power-law + Gaussian line" model, and we used the best-fit slopes to construct
their sample distribution function. We used the median of the distributions, and
the mean of the best-fit slopes which are above the 80th percentile of the
distributions, to estimate the mean spectral slope of the objects (the latter
estimate is more appropriate in the case when the energy spectra of the sources
are significantly affected by absorption and/or reflection effects). We also
used results from literature to estimate the average spectral slope of three
more objects. The fourteen AGN that we consider in this work are nearby, X--ray
bright objects, whose X--ray light curves have been studied in the past. Their
PSDs have been accurately estimated, and characteristic ``break frequencies"
have been detected in them. 

When we combine the mean spectral slope estimates with the $\nu_{\rm br}$
estimates listed in UM05, we find that: {\it objects with steeper mean energy
spectra have shorter characteristic time scales as well}. This is the first time
that such a ``spectral -- timing" correlation is detected in AGN. The results we
reported in Section 4.3 suggest that this spectral-timing relation in AGN can be
parametrised as follows: $\Gamma\approx 1.3\nu_{\rm norm}^{0.07}$. 

\subsection{The spectral-timing correlation in AGN}

The easiest way to explain this correlation is to assume that, for an AGN with a
given BH mass, the accretion rate determines both  the PSD characteristic
frequencies (this has already been shown by M06) and their energy spectral shape
as well: the higher the accretion rate, the steeper the mean energy spectrum
will be. Such a positive $\Gamma-\dot{\rm m}_{\rm E}$ relation in AGN has
already been suggested/shown in the past (e.g. Porquet \etal\ 2004;  Bian 2005;
Shemmer \etal\ 2006; Saez \etal\ 2008). A positive $\Gamma-\dot{\rm m}_{\rm E}$
relation has also been detected recently in 7 GBHs by Wu \& Gu (2008), when they
accrete at a rate higher than $\sim 1\%$ of the Eddington limit. Our results are
consistent with the results reported in these papers.

Using the above results and the M06 results, we calculate that $\nu_{\rm
norm}($Hz$\times$M$_{\odot})\approx 3000\dot{\rm m}_{\rm E}$. When we  replace
$\nu_{\rm norm}$ in the relation of  $\Gamma\approx 1.3\nu_{\rm norm}^{0.07}$ we
found in this work, we find that: $\Gamma \approx 2.3\dot{\rm m}_{\rm
E}^{0.07}$.  This formula between spectral slope and accretion rate could be
helpful in deriving a rough estimate for the  accretion rate of the  distant AGN
that are detected in deep X-ray surveys, assuming that their variability
properties are similar to the properties of nearby AGN (this is supported by
recent studies, see e.g. Papadakis \etal\ 2008).  

In the context of thermal Comptonisation models, $\dot{\rm m}_{\rm E}$ can
affect the spectral slope $\Gamma$ as it controls the strength of the soft disc
photons, hence the cooling of the thermal  plasma in the X-ray emitting corona.
The greater the cooling by seed photons incident on the plasma, the softer the
resulting X-ray power-law spectra are. Indeed, we found that Comptonisation
models are consistent with the $\Gamma - \dot{\rm m}_{\rm E}$  relation that our
results imply,  but  only if $L_{\rm s}/L_{\rm diss}\propto \dot{\rm m}_{\rm
E}$.  

\subsection{The comparison with Cyg X-1}

Both P03 and A06 have used data from \rxte\ observations that lasted for $\sim
1-2$ ks, and were performed every $\sim 5-10$ days over a period of many years.
If ``time scales" scale with BH mass in accreting objects, then $1-2$ ksec in
Cyg X-1 should correspond to a period of at least $\sim 3-6$ years in objects
with M$_{\rm BH}> 10^6$ times the mass of the black hole in Cyg X-1  (like most
AGN in our sample). This suggests that each one of the AGN points in Fig.~3
corresponds to just one of the Cyg X-1 points plotted in the same figure. We
found that the AGN and Cyg X-1 ``$\Gamma - \nu_{\rm norm}$" relations are
similar but not the same. In Section 5 we discussed briefly some implications of
this result. In the paragraphs below we discuss the implications of our results
in more detail. 

The main aim of the  discussion below is to investigate the constrains that
the AGN spectral-timing relation, and its comparison with the similar relation
in Cyg X-1, impose on thermal Comptonisation models, based on the particular
assumption that both the spectral and timing properties of accreting systems are
driven by accretion rates variations. We point out that other interpretations
are also possible; see for example Kylafis \etal\ (2008) for an alternative
explanation of the Cyg X-1 spectral-timing relation, which does not assume that
X--rays are produced by thermal Comptonisation. We also point out that, given
the small number of objects in our sample, and the unavoidable uncertainty in
the derived parameters of the AGN spectral-timing relation, the values of the
various parameters in the equations below are somehow uncertain.

As we showed in Section 6.1, the AGN spectral-timing relation and the  M06
results imply that $\Gamma_{\rm AGN} \approx 2.3  \dot{m}_{\rm E}^{0.07}$.  If
X--rays in AGN  are produced by thermal Comptonisation, we expect\footnote{The
discussion in these paragraphs is based on the assumption of an outflowing
corona, hence we adopt the results of Malzac \etal, 2001. Similar conclusions
can also be drawn if we assume a static corona.} that $ \Gamma_{\rm AGN} = 2.2
(L_{\rm s}/L_{\rm diss})^{0.07}$.  Therefore, the observations [i.e. 
$\Gamma_{\rm AGN} \approx 2.3  \dot{m}_{\rm E}^{0.07}$] are consistent with the
thermal Comptonisation model predictions [$ \Gamma_{\rm AGN} = 2.2 (L_{\rm
s}/L_{\rm diss})^{0.07}$], only if

\begin{equation}
(L_{\rm s}/L_{\rm diss})\approx 2 \dot{m}_{\rm E}.
\end{equation}

\noindent An obvious implication of this result is that, if a certain fraction,
say $f_{\rm s}$, of the total accretion power, $P_{\rm tot}$
$(=\eta\dot{m}_{\rm E}c^2,$ where $\eta$  is the efficiency of the accretion
process), is converted to disc luminosity (i.e. $L_{\rm s}=f_{\rm s}P_{\rm
tot}$), while  $L_{\rm diss}=f_{\rm diss}P_{\rm tot}$, then the ratio $(f_{\rm
s}/f_{\rm diss})$ should not remain constant for a given object, but should
rather increase with increasing accretion rate. 

Suppose that X--rays from Cyg X-1 in its low/hard state (LH) are also produced
by thermal Comptonisation. In this case,  thermal Comptonisation models predict
that $\Gamma_{\rm CygX-1} = 2.2 (L_{\rm s}/L_{\rm diss})^{0.13}$, or 

\begin{equation}
\Gamma_{\rm CygX-1}\approx  2.4 \dot{m}_{\rm E}^{0.13},
\end{equation}

\noindent if we accept that equation (1) holds in this case as well. According
to K\"{o}rding \etal\ (2007), the normalisation of the $\nu_{\rm norm}$ vs
$\dot{m}_{\rm E}$ relation for the hard-state GBHs is $\sim 8$ times smaller
than the normalisation in the case of the AGN in our sample. So, if  $\nu_{\rm
norm, Cyg X-1/LS}\approx 375\dot{m}_{\rm E}$, as opposed to $\nu_{\rm norm,
AGN}\approx 3000\dot{m}_{\rm E}$, then  $\dot{m}_{\rm E}\approx 3\times
10^{-3}\nu_{\rm norm, Cyg X-1/LS}.$ We can now substitute  $\dot{m}_{\rm E}$ in
equation (2) to determine the $\Gamma -\nu_{\rm norm}$ relation for Cyg X-1 in
LH state:

\begin{equation}
 \Gamma_{\rm CygX-1/LS}\approx  1.1 \nu_{\rm norm}^{0.13}.
\end{equation}

\noindent The dashed line in the top panel of  Fig.~3 indicates this relation.
The agreement between the {\it predicted} $\Gamma-\nu_{\rm norm}$ relation  and
the Cyg X-1 data is rather good up to $\nu_{\rm norm}\sim 70$. The discussion so
far suggests the following picture:

a) X-rays from the AGN we studied are produced by thermal Comptonisation.  The
$\Gamma-\nu_{\rm norm}$ relation we observe is consistent with the predictions
of thermal Comptonisation models but only if the $(L_{\rm s}/L_{\rm diss})$
ratio, and hence the $(f_{\rm s}/f_{\rm diss})$ ratio as well, increase
proportionally with accretion rate.

b) X-rays in Cyg X-1 are also produced by thermal Comptonisation. Taking into
account the fact that the normalisation of the $\nu_{\rm norm}-\dot{m}_{\rm E}$
relation is $\sim 8$ times smaller in Cyg X-1 than in the AGN in our sample
(K\"ording \etal\ 2007), the {\it predicted} $\Gamma-\nu_{\rm norm}$ relation
agrees well with the Cyg X-1 data up to $\nu_{\rm norm}\approx 70$.

c) In the case when $\Gamma_{\rm AGN}=\Gamma_{\rm Cyg X-1}$, we expect that
($L_{\rm s}/L_{\rm diss})_{\rm Cyg X-1}^{0.13}=(L_{\rm s}/L_{\rm diss})_{\rm
AGN}^{0.07}$, and $\dot{m}_{\rm AGN,E}^{0.07}=\dot{m}_{\rm Cyg X-1,E}^{0.13}$
(using equation 1). Consequently, $\dot{m}_{\rm E,AGN}=\dot{m}_{\rm E,Cyg
X-1}^{1.9}$ and, since  $\dot{m}_{\rm E}<1$, AGN should operate on a lower
accretion rate than Cyg X-1 when the spectral slope is the same in both systems.
Furthermore, since $\Gamma_{\rm AGN}\approx 1.3 \nu_{\rm norm,AGN}^{0.07}$  and
$\Gamma_{\rm Cyg X-1}\approx 1.1 \nu_{\rm norm,Cyg X-1}^{0.13}$ (when 
$\Gamma\la 2.1-2.2$), then  $\Gamma_{\rm AGN}=\Gamma_{\rm Cyg X-1}$ implies that
$1.3 \nu_{\rm norm,AGN}^{0.07}=1.1 \nu_{\rm norm,Cyg X-1}^{0.13}$, and 
$\nu_{\rm norm,Cyg X-1}\approx 3.5\nu_{\rm norm,AGN}^{0.55}$. Therefore, when
the spectral slope is the same (and less than $\sim 2.1-2.2$) in AGN and Cyg
X-1, the former should operate at a lower accretion rate but their 
characteristic time scales should be shorter than those in Cyg X-1 (when
normalised to the respective BH mass), because the normalisation of the  AGN
$\dot{m}-\nu_{\rm norm}$ relation is significantly larger than the normalisation
of the respective Cyg X-1 relation in LH state.

The vertical dot-dashed line in the top panel of Fig.~3  indicates the value
$\nu_{\rm norm}=70$.  For Cyg X-1,  this normalised frequency corresponds to 
$\nu_{1}=1$ Hz and $\nu_2=5$  Hz for the centroid frequency of the  low and
higher frequency Lorentzians, respectively. As $\nu_{\rm norm}$ (i.e. the
accretion rate) increases even more, then i)  the ratio $\nu_{1}/\nu_2$
increases as well (see Fig.~2 in A06), ii) the contribution of the Lorentzians
to the root mean square variability amplitude decreases, and iii) the ``bending"
power-law component in the PSD appears and its contribution to the  source
variability amplitude increases (see Fig.~7 in A06), i.e. the Cyg X-1 power
spectrum changes from a hard to a soft-state shape. At $\nu_{\rm norm}\approx
70$, the average {\it observed} spectral slope of Cyg X-1 is $\sim 2.1$ (see
Fig.~3), while  at higher frequencies the spectral slope is steeper.
Consequently, the region defined by  $\Gamma<2.1$ and $\nu_ {\rm norm}<70$
corresponds to the ``hard state region" for Cyg X-1 in the ``spectral-timing
plane" shown in Fig.~3. 

The discrepancy between the predicted spectral-timing relation and the Cyg X-1
data when $\Gamma\ma 2.1$ and $\nu_{\rm norm}\ma 70$ cannot be explained by the
fact that the normalisation of the  $\nu_{\rm norm} - \dot{m}_{\rm E}$ relation
increases by a factor of $\sim 8$ in the high/soft state.  If that were the
case,  the  spectral-timing relation in this state should be similar to the one
defined by equation (3), but with a {\it smaller} normalisation (opposite to
what we observe).

One possibility is that the  $(L_{\rm s}/L_{\rm diss})-\dot{m}_{E}$ relation
(equation 1) changes in the high/soft state. However, in this case we would have
to accept that the X--ray source does  {\it not}  operate in the same way in AGN
and Cyg X-1: when  $\nu_{\rm norm}>70$, both Cyg X-1 and the AGN in our sample
follow the same $\nu_{\rm norm} - \dot{m}_{\rm E}$ relation (M06, K\"ording
\etal\ 2007). Therefore, as long as $\nu_{\rm norm}>70$, a given  $\nu_{\rm
norm}$ value implies the  same accretion rate in both systems. The fact that the
AGN spectral-timing relation is valid up to $\nu_{\rm norm}\approx 1000$  should
then imply that, for the same accretion rate,  the  ``$(L_{\rm s}/L_{\rm diss})
- $ accretion rate" relation is different in AGN and Cyg X-1. 

In Cyg X-1, $\Gamma\approx 2.1$  implies that $2.2 (L_{\rm s}/L_{\rm
diss})^{0.13}\approx 2.1$, and hence  $(L_{\rm s}/L_{\rm diss})\approx 0.7$.
Another possible explanation then for the discrepancy between the Cyg X-1 data
and the predicted spectral-timing relation above $\nu_{\rm norm}=70$ is the
following: equation (1) holds until $(L_{\rm s}/L_{\rm diss})\approx 0.7$, at
which point the hot corona is significantly cooled down, and the thermal X--ray
emission component is weak. It is possible then that at high accretion rates  a
separate, possibly non-thermal, X--ray component emerges, and dominates the
X-ray emission in the soft state. If that is the case, the $\Gamma- L_{\rm
s}/L_{\rm diss}$ relation we have assumed above is not valid, hence the
predicted spectral-timing relation does not fit the data any more. 

Even if the picture drawn above is correct, there are important issues regarding
the relation between AGN and GBH states which still remain unresolved.  In
particular, the answer to the question whether the AGN in our sample are 
``soft" or ``hard" state systems is far from clear. There are indications that
they are analogous to Cyg X-1 in its soft state. For example, the radio emission
in Cyg X-1 in its LH state is enhanced. On the other hand, Panessa \etal\ (2007)
have shown that, for the same $L_{\rm X}/L_{\rm E}$ ratio, the radio luminosity
of  Seyfert galaxies is $\sim 8-10$ times lower than the radio luminosity of
hard-state GBHs, even when the BH mass difference is properly taken into account
(7 of the objects in our sample are also included in their sample). Furthermore,
the AGN in our sample follow the soft-state ``characteristic time scale --
accretion relation" in GBHs, and  a ``bending" power-law is the dominant
component in the power spectrum of those objects with good enough light curves
to accurately study their PSD (e.g. NGC~4051, M$^{\rm c}$Hardy \etal\ 2004;
NGC~3227 and  NGC~5506, UM05; MCG~6-30-15, M$^{\rm c}$Hardy \etal\ 2005;  and
perhaps NGC~3783, Summons  \etal\ 2007), implying a soft-state like PSD in these
objects. High quality light curves for low accretion rate AGN are necessary to
investigate whether any AGN with hard-state like power spectra exist or not. 

However, although the radio emission strength and the timing properties of many
objects in our sample are soft-state like, their spectral properties are {\it
not}, as the average spectral slope is smaller than $2.1$ in most cases. If
indeed the {\it spectral} hard-to-soft state transition corresponds to $(L_{\rm
s}/L_{\rm diss})\sim 0.7$, at which point the thermal corona emission is weak,
and a different component dominates the X--ray emission, then we should expect
this transition to happen when $\Gamma\ma 2.15$ in AGN. Given the AGN
$\Gamma-\nu_{\rm norm}$ relation, this slope corresponds to $\nu_{\rm
norm}\approx 2500$. At even higher normalised  frequencies, we should then
expect the AGN spectral-timing relation to break (like in Cyg X-1 at $\nu_{\rm
norm}\approx 70$). Obviously, more data are necessary to confirm that this is
indeed the case in AGN. 

The data so far suggest that while in AGN the {\it timing} properties transition
from hard-to-soft state happens at least as low as  $\nu_{\rm norm}\approx 100$,
the {\it spectral} properties transition  should happen at much higher accretion
rates. This is opposite to what we observe in Cyg X-1, where both the spectral
and timing properties change from the hard to soft state, at the same accretion
rate (indicated by the value $\nu_{\rm norm}\approx 70)$. Perhaps the timing
properties in accreting objects are determined by accretion disc variations
only, and the hard-to-soft state transition materialises at a certain accretion
rate, irrespective of whether the soft disc luminosity is strong enough for
$(L_{\rm s}/L_{\rm diss})> 0.7$, i.e. strong enough to cool down the hot corona.
In this case, we would expect the AGN hard-to-soft timing properties transition
to appear at $\nu_{\rm norm}\approx 70$ as well. Due to the cooler disc
temperature though, the AGN spectral soft-to-hard state transition happens at
higher accretion rates (i.e. at a higher $\nu_{\rm norm}$ value in the
spectral-timing plane of Fig.~3). Further progress in understanding the relation
between AGN and GBHs can be made when we know how the accretion rate determines
the characteristic frequency in accreting compact objects (assuming that it is
just the accretion rate that determines $\nu_{\rm norm}$ in these systems).

\begin{acknowledgements} 

We would like to thank the referee, P. O. Petrucci, for valuable comments which
helped us to improve the paper significantly. IEP and MS acknowledge support by
the EU grant MTKD-CT-2006-039965. MS also acknowledges support by the the Polish
grant N20301132/1518 from Ministry of Science and Higher Education.

\end{acknowledgements}

\end{document}